\documentclass[letterpaper, 10 pt, conference]{ieeeconf}  

\usepackage{amsmath,epsfig}
\usepackage{balance}
\usepackage{cite}
\usepackage{graphicx}
\usepackage{lipsum}
\usepackage{url}
\graphicspath{{Figures/}}
\usepackage{caption}
\usepackage{subfig} 
\usepackage{color}
\usepackage{balance}
\usepackage{float}
\usepackage{multirow}
\usepackage{float}
\usepackage{breqn}
\usepackage{tabularx}

\IEEEoverridecommandlockouts                              

\title{\LARGE \bf
 Predicting Stroke from Electronic Health Records 
}

\author{Chidozie Shamrock Nwosu$^{\star}$$^{1}$, Soumyabrata Dev$^{\star}$$^{2}$, Peru Bhardwaj$^{3}$, \\ Bharadwaj Veeravalli$^{4}$, and Deepu John $^{5}$
\thanks{$^{\star}$ Authors contributed equally.}
\thanks{$^{1}$C.\ S.\ Nwosu is with the School of Computing, National College of Ireland, Ireland.
        {\tt\small x17161916@student.ncirl.ie}}%
\thanks{$^{2}$S.\ Dev is with the ADAPT SFI Research Centre, Trinity College Dublin, Ireland.
        {\tt\small soumyabrata.dev@adaptcentre.ie}}%
\thanks{$^{3}$P.\ Bhardwaj is with the ADAPT SFI Research Centre, Trinity College Dublin, Ireland.
        {\tt\small peru.bhardwaj@adaptcentre.ie}}%
\thanks{$^{4}$B.\  Veeravalli is with the School of Electrical and Computer Engineering, National University of Singapore, Singapore.
        {\tt\small elebv@nus.edu.sg}}%
\thanks{$^{5}$D.\ John is with the School of Electrical and Electronic Engineering, University College Dublin, Ireland.
        {\tt\small deepu.john@ucd.ie}}%
}

\begin{document}

\maketitle
\thispagestyle{empty}
\pagestyle{empty}


\begin{abstract}

Studies have identified various risk factors associated with the onset of stroke in an individual. Data mining techniques have been used to predict the occurrence of stroke based on these factors by using patients' medical records. However, there has been limited use of electronic health records to study the inter-dependency of different risk factors of stroke. In this paper, we perform an analysis of patients' electronic health records to identify the impact of risk factors on stroke prediction. We also provide benchmark performance of the state-of-art machine learning algorithms for predicting stroke using electronic health records.

\end{abstract}

\section{INTRODUCTION}
\label{sec:intro}

The field of medicine has witnessed great improvement because of technological advancements. 
On one hand, it has become easier to collect a range of healthcare data due to low-cost wearable devices~\cite{veeravalli2017real}. On the other hand, insights acquired from mining clinical data have proved useful for decision making in improving healthcare and reducing healthcare costs~\cite{sanyal2018algorithms,yoo2012data}. Data mining techniques are used to identify risk factors associated with the onset of a disease~\cite{deepu2015low, deepu20182}. Of particular interest to us, is the identification of risk factors associated with stroke. 

Studies in healthcare like~\cite{meschia2014guidelines,harmsen2006long,schneider2003trends} have identified high blood pressure, diabetes and heart disease as major risk factors responsible for stroke attack in an individual. Several machine learning algorithms have also been proposed to use these risk factors for predicting stroke occurrence~\cite{jeena2016stroke,hanifa2010stroke}. However, a systematic analysis of the risk factors is missing. We attempt to bridge this gap by analyzing the correlation between different risk factors for stroke prediction. We employ Principal Component Analysis to determine if the feature space for predictive modelling can be reduced by removing the input features that are highly correlated. We also compare and benchmark the performance of state-of-art machine learning algorithms for the task of classifying a patient's health record as suffered from stroke or not. Hence, the main contributions of this paper are twofold -- (a) we perform a systematic analysis of risk factors for stroke prediction; (b) we also provide a benchmark performance for stroke prediction using the state-of-art machine learning algorithms. Finally, in the spirit of reproducible research, we have made available the source code of all simulations used in this paper\footnote{\url{https://github.com/Soumyabrata/analyzing-stroke}}.

\subsection{Related Work}
There are several works in literature that use machine learning techniques on electronic health records to predict the probability of stroke occurrence. Jeena et al.\ \cite{jeena2016stroke} identified that there is a direct relationship between the total count of risk factors and the probability of stroke occurrence. A regression approach was recommended to statistically test the association between a risk factor and its effect. Hanifa and Raja~\cite{hanifa2010stroke} achieved an improved accuracy for predicting stroke risk using radial basis function and polynomial functions applied in a non-linear support vector classification model. At the same time, studies have indicated that redundant attributes and/or totally irrelevant attributes to a class should be identified and removed before the use of a classification algorithm~\cite{yoo2012data}. Systematic analysis of input features has been performed for modelling the response variable in areas other than healthcare -- color analysis of ground-based sky/cloud images~\cite{dev2014systematic}, weather recordings for rainfall detection~\cite{manandhar2018systematic} etc.

\subsection{Electronic Health Records Dataset}
We use a dataset of electronic health records released by McKinsey \& Company as a part of their healthcare hackathon challenge\footnote{\url{https://datahack.analyticsvidhya.com/contest/mckinsey-analytics-online-hackathon/}}. The dataset is available from Kaggle\footnote{\url{https://inclass.kaggle.com/asaumya/healthcare-dataset-stroke-data}}, a public data repository for datasets. 
The dataset contains $29072$ patient's information with $12$ attributes. The $11$ input attributes are as follows: patient identifier, gender, age, binary status $1/0$ if the patient is suffering from hypertension or not, binary status $1/0$ if the patient is suffering from heart disease or not, marital status, type of occupation, type of residence (urban/ rural), average glucose level, body mass index, and patient's smoking status. The $12^{th}$ attribute is the binary response variable $1/0$ indicating if the patient has suffered stroke or not. Ethical requirements were processed by the publishers of the dataset in compliance with data protection laws. In the subsequent sections of this paper, we consider the $10$ patient attributes (excluding the patient identifier) as the input variables, and the binary response variable of stroke as the target variable for a predictive model.

\section{Principal Component Analysis}
\label{sec:pca}

In this section, we analyze the variance in the dataset using Principal Component Analysis (PCA). We perform PCA on the electronic health records dataset using the $10$ patient attributes as variables. The dataset is represented by a variable matrix $\mathbf{X}$ of dimension $m \times n$, where $m$ is the total number of input variables, and $n$ is the total number of patient records. In our case, the values are $m=10$ and $n=29072$. The individual features $f_{1-10}$ are extracted from $\mathbf{X}$, and reshaped into corresponding column vectors $\mathbf{\widetilde{v}}_j \in {\rm I\!R}^{mn \times 1}$ where $j=1,2,..,10$. These input vectors $\mathbf{\widetilde{v}}_j$ are stacked to form the matrix  $\hat{\textbf{X}} \in {\rm I\!R}^{mn \times 10}$:

\begin{equation}
\label{eq:eq1}
\hat{\textbf{X}}=[\mathbf{\widetilde{v}}_1, \mathbf{\widetilde{v}}_2,\mathbf{\widetilde{v}}_3,\ldots,\mathbf{\widetilde{v}}_{10}].
\end{equation}

$\hat{\textbf{X}}$ is normalized with corresponding mean $\bar{v_{j}}$ and standard deviation $\sigma_{v_{j}}$ of the individual features. This normalised matrix $\ddot{\mathbf{X}}$ is represented as:

\begin{equation}
\label{eq:eq3}
\ddot{\mathbf{X}}= \left[\frac{\widetilde{\mathbf{v}_{1}}-\bar{v_{1}}}{\sigma_{v_{1}}}, \frac{\widetilde{\mathbf{v}_{2}}-\bar{v_{2}}}{\sigma_{v_{2}}},..,\frac{\widetilde{\mathbf{v}_{j}}-\bar{v_{j}}}{\sigma_{v_{j}}},..,\frac{\widetilde{\mathbf{v}_{10}}-\bar{v_{10}}}{\sigma_{v_{10}}}\right].
\end{equation}

Subsequently, the covariance matrix of $\ddot{\mathbf{X}}$ is computed. The eigenvalue decomposition of the covariance matrix then yields the eigenvalues and the eigenvectors. The eigenvectors define the new orthogonal axes called principal components which act as summaries of the features of the dataset. PCA can be used to reduce the feature space for predictive modelling if the first few components capture most of the variance in the data.

\subsection{Variance explained by principal components}
\begin{figure}[htb]
  \begin{center}
    \includegraphics[width=0.40\textwidth]{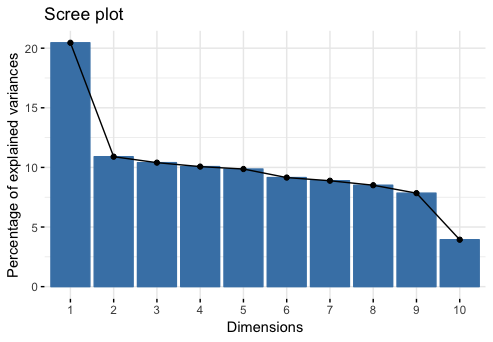}
  \end{center}
  \caption{Percentage of variance explained by different principal components.}
  \label{fig:scree}
\end{figure}

Figure~\ref{fig:scree} shows the scree plot for our study, indicating the percentage of variance in the dataset explained by the different principal components. 
We understand that the first $9$ principal components explain $96.06\%$ variance in the dataset. Thus, reducing the feature space for predictive modelling to $9$ features will result in only about $4\%$ loss of explained variance. However, this should not be done because the input features are not highly correlated. This is because the first two principal components can capture only $31.4\%$ variance and there is very little difference in the percentage of variance explained by each principal component. Hence, all features should be used for predictive modelling.

\subsection{Relation between principal components with patient attributes}

\begin{figure}[htb]
  \begin{center}
    \includegraphics[width=0.45\textwidth]{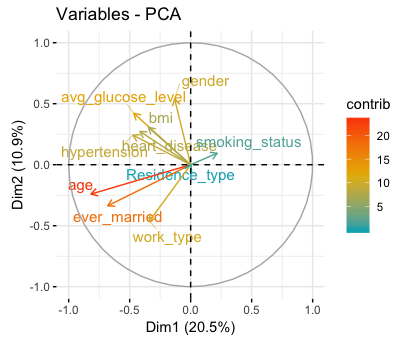}
  \end{center}
  \caption{Biplot representation of the actual patient attributes projected on the first two principal components.}
  \label{fig:biplot}
\end{figure}

Figure~\ref{fig:biplot} is a biplot representation which shows the correlation of patient attributes with the first two principal components. The two components are represented by two orthogonal axes. Each patient attribute is represented by a vector and the length of vector shows the importance of the corresponding attribute in interpreting the principal components. If the dataset were perfectly represented by the first two principal components, vector heads of all variables would be positioned on the circle, which is called the circle of correlations. Since all vectors are positioned inside the circle of correlations, this means that more than two components are needed to represent the data perfectly.

\begin{figure*}[htb]
  \begin{center}
    \subfloat[]{\includegraphics[width=0.45\textwidth]{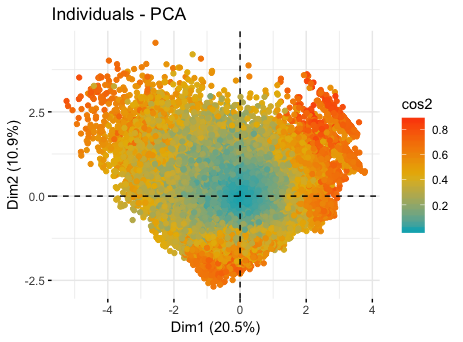}}
    \subfloat[]{\includegraphics[width=0.45\textwidth]{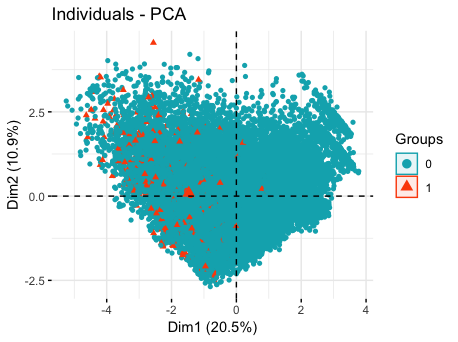}}
  \end{center}
  \caption{Sub-space representation of the different observations in the electronic health records transformed on the reference of the first two principal components. Each observation is color coded based on (a) \emph{cos2} measure indicating the importance of principal components on it; and (b) the binary $0/1$ response variables of the stroke condition.}
  \label{fig:pca-individual}
\end{figure*}

We can see from Figure~\ref{fig:biplot} that patient's residence type has no contribution to the first two principal components. The patient's age and marital status have the highest contribution to the first two principal components.
The first principal component (horizontal axis) contrasts a patient's smoking status with the remaining attributes. This indicates that older married patients with high glucose level, hypertension and heart disease do not smoke.
We can also see that the second component (vertical axis) contrasts a patient's average glucose level, body mass index, gender, hypertension and heart disease status with their age, marital status and work type. This indicates that patient's with high glucose level, hypertension and heart disease have less age and are not married. Hence, it appears that the first principal component shows a patient's smoking status and the second principal component separates a patient's medical characteristics with their lifestyle.

\subsection{Relation of Principal Components with individual records}

Figure~\ref{fig:pca-individual} shows the projection of each patient record in the reference of first two principal components. Each patient record or observation is represented by a point in the sub-space.
Figure~\ref{fig:pca-individual}(a) shows the importance of the first two principal components in representing the individual medical health records. The squared cosine indicates the contribution of the component to the squared distance of an observation from the origin. Hence, observations that are well represented by the first two principal components are positioned farther away from the origin. These are shown in red. The principal components are not important in representing the observations that are closer to the origin, which are shown in blue. The plot indicates that the first two principal components contribute to the representation of only a part of the observations. A number of observations that are closer to the origin cannot be represented by the first two principal components.

Figure~\ref{fig:pca-individual}(b) shows the binary response variable indicating the stroke condition for each observation. We observe that the observations with $0$ label (no stroke condition) and $1$ label (suffered stroke) are not well separated into distinct clusters in the sub-space of first two principal components. This indicates that more features are required to separate the labels in a higher-dimension space.

\subsection{Discussion}

In this section, we provided an analysis of our results from PCA. We found that the first two principal components can capture only $31.4\%$ of the total variance in the data, which indicates that the patient attributes are not highly correlated. We studied the contribution of different patient attributes to the first two principal components and found that the two components do not represent the health records data perfectly. We also looked at the contribution of the first two principal components in the representation of individual health records and found that some health records can be represented by the two components, but some can not. Our analysis shows that all patient attributes should be used as features for predictive modelling of stroke occurrence.

\section{Stroke Prediction}
\label{sec:prediction}
In this section, we perform a comparative analysis of three popular classification algorithms -- neural network, decision tree and random forest on our dataset of medical records. 

Based on the analysis in previous section, we use all the $10$ input variables for training our binary classification task of stroke prediction. Our dataset of electronic medical records is highly unbalanced in nature. Out of $29072$ medical records, only $548$ records belong to patients who suffered from stroke condition, the rest $28524$ records belong to patients with no stroke condition. This poses a difficult problem in training any machine-learning based model, as the dataset is highly unbalanced in nature. Therefore, we employ a random down-sampling technique to reduce the adverse effect of unbalanced dataset. We refer the $548$ observations as minority class, while remaining $28524$ observations as majority class. We create a balanced dataset consisting of $548$ minority- and $548$ majority- observations in the following order: we choose all the available observations of minority cases and randomly choose $548$ observations from $28524$ majority case observations. The balanced dataset has $1096$ observations. We use $70\%$ of the dataset for training the machine learning models and $30\%$ of the dataset for testing their performance. We use the classification accuracy as the evaluation metric to measure the performance of the machine learning models.

In order to remove sampling bias, we perform $1000$ such random downsampling experiments. Table~\ref{table:compare} reports the average classification accuracy for the three benchmarking approaches. We observe that the performance of decision tree and random forest are similar. We obtain the best accuracy result of $75.02\%$ from the feed-forward multi-layer perceptron model. This makes sense as neural networks work better with multivariate input variables, and can handle noisy data. 

\begin{table}[t]
\centering
\normalsize
\caption{Average binary classification accuracy of neural network, decision tree and random forest over 1000 experiments on our dataset of electronic health records.}
\begin{tabular}{ll}
\hline
\textbf{Approach} & \textbf{Average accuracy} \\
 & \textbf{(over 1000 experiments)} \\
\hline 
Decision Tree & 74.31\%    \\
Random Forest & 74.53\%   \\
Neural Network  & 75.02\%     \\
\hline
\end{tabular}
\label{table:compare}
\end{table}

\begin{figure}[htb]
  \begin{center}
    \includegraphics[width=0.5\textwidth]{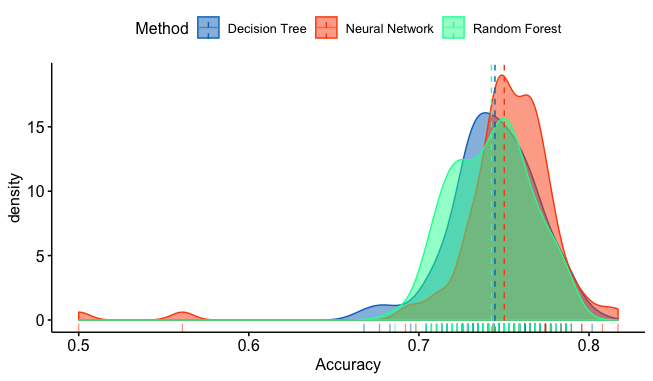}
  \end{center}
  \caption{Histogram distribution of binary classification accuracy of benchmarking algorithms over $1000$ experiments.}
  \label{fig:accurate-dist}
\end{figure}

As a final comparison, we compute the density distribution of classification accuracy for the benchmarking methods, across all the $1000$ experiments. Figure~\ref{fig:accurate-dist} shows this distribution. For example, we can see that the neural network has the classification accuracy of $75.02\%$ for more than $15\%$ experiments. We observe that all algorithms significantly overlap with each other around their corresponding mean values. Hence, the algorithms have similar classification performance.

\section{Conclusion and Future Work}
\label{sec:conc}
In this paper, we provide a systematic analysis of risk factors for stroke prediction. These risk factors are represented as patient attributes in electronic health records. We use principal component analysis to analyze the sub-space representation of $10$ attributes into two principal components. The analysis shows that since the patient attributes are not highly correlated, the feature space for predictive modelling cannot be reduced significantly without a significant loss of information. Hence, we use all patient attributes as input features for stroke prediction. We compare the performance of different state-of-art classification algorithms, namely decision trees, random forests and neural networks for predicting stroke. We find that the multi-layer perceptron model has the best performance with an accuracy of $75.02\%$. 

There are multiple directions for further work in future. First, we plan to study the effect of the use of a subset of features on the accuracy of classification algorithms. Secondly, we plan to integrate the electronic records dataset with background knowledge on different diseases and drugs from the publicly available Linked Open Data (LOD) cloud, as demonstrated in \cite{tilahun2014design}. This can be done by uplifting the dataset into Linked Data and generating interlinks to the LOD cloud. This interlinking process is not trivial~\cite{bhardwaj2018overlooked} and the generated interlinks need to be maintained~\cite{singh2018delta} but the integration of background knowledge can improve the performance of classification algorithms for stroke prediction.

\balance


\end{document}